\documentclass[aps,prb,reprint,twocolumn,superscriptaddress,showpacs,floatfix]{revtex4-1}

\usepackage{hyperref}
\usepackage{graphicx}
\usepackage{amsmath}
\usepackage{amssymb}
\usepackage{bbm}
\usepackage{enumitem}
\begin{document}

\title{Testing self-energy embedding theory in combination with GW}
\author{Tran Nguyen Lan}
\altaffiliation{On leave from: Ho Chi Minh City Institute of Physics, VAST, Ho Chi Minh City, Vietnam}
\affiliation{Department of Chemistry, University of Michigan, Ann Arbor, Michigan 48109, USA}
\affiliation{Department of Physics, University of Michigan, Ann Arbor, Michigan 48109, USA}
\author{Avijit Shee}
\affiliation{Department of Chemistry, University of Michigan, Ann Arbor, Michigan 48109, USA}
\author{Jia Li}
\affiliation{Department of Physics, University of Michigan, Ann Arbor, Michigan 48109, USA}
\author{Emanuel Gull}
\affiliation{Department of Physics, University of Michigan, Ann Arbor, Michigan 48109, USA}
\author{Dominika Zgid}
\affiliation{Department of Chemistry, University of Michigan, Ann Arbor, Michigan 48109, USA}
\date{\today}
\begin{abstract}
We present a theoretical framework and implementation details for self-energy embedding theory (SEET) with the GW approximation for the treatment of weakly correlated degrees of freedom and configuration interactions solver for handing the strongly correlated degrees. On a series of molecular examples, for which the exact results are known within a given basis, we demonstrate that SEET(CI/GW) is a systematically improvable and well controlled method capable of giving accurate results and well behaved causal self-energies and Green's functions. We compare the theoretical framework of SEET(CI/GW) to that of GW+DMFT and comment on differences between these to approaches that aim to treat both strongly and weakly correlated simultaneously.
\end{abstract}
\maketitle
%\noindent
%\textcolor{cyan}{TNL color is cyan}\\
%\textcolor{green}{AS color is green}\\
%\textcolor{magenta}{JL color is magenta}\\
%\textcolor{red}{EG color is red}\\
%\textcolor{blue}{DZ is blue}\\

\section{Introduction}

The quantitative simulation of realistic correlated solids and molecules requires the accurate description of a large number of degrees of freedom. Only some of these are strongly correlated. A successful approach has therefore been the combination of weak correlation methods, such as the GW method or the density functional theory in the local density approximation (LDA), with non-perturbative methods for low-energy effective models, such as the dynamical mean field approximation.

The polynomial scaling of the weak correlation method then allows one to treat large systems, while the dynamical mean field approximation replaces the intractable lattice problem with an impurity problem coupled to a self-consistently adjusted bath, which is numerically solvable.

The combination of band structure methods with dynamical mean field theory (DMFT)~\cite{Georges92,Georges96} has enjoyed great success, in particular when applied to materials with $d$ and $f$ shells. However, ambiguities at the interface of the weak- and strong-correlation methods, related in particular to the choice of ``double counting'' and the impurity interaction parameters in LDA+DMFT~\cite{savrasov_RevModPhys06}, and to the proper numerical treatment of general four-fermion screened interaction terms in GW+DMFT~\cite{Biermann03,Biermann05,Casula16}, have hampered progress.
Moreover, the assumption that strong correlations are spatially localized, which is the premise of the dynamical mean field approximation, may not be valid in real compounds.

Recently we introduced an approximation scheme, the self-energy embedding theory (SEET)\cite{Zgid15,Tran15,Tran16,Tran17,zgid_njp17,simons_benchmark2} that does not suffer from these limitations. First, due to the diagrammatic nature of the method, no double counting problem arises. 
Second, the absence of frequency-dependent interactions in the strongly correlated part means that standard multi-orbital impurity solvers can be employed and no ambiguity in how the Coulomb interactions are handled exist. 
Finally, the method does not rely on an a priori determination of the correlated subspace and in particular does not assume that strong correlations are local, but rather introduces a small control parameter that can be used to adaptively choose orbitals and systematically converge to the exact solution.

In a series of previous papers \cite{Zgid15,Tran15,Tran16,Tran17,zgid_njp17,simons_benchmark2}, we studied the behavior of SEET where the weak-correlation method employed was the second order perturbation theory (GF2). In the present paper, we show how SEET performs when GF2 is replaced by the GW approximation.
QM/QM embedding methods for realistic systems are not well studied, and prior work on non-perturbative strong correlation methods for small systems based on diagrammatic theory have shown problems ranging from convergence to the ``wrong'' fixed point~\cite{Kozik15} to causality violations~\cite{Haule17_h2}. We show here that no such problems are observed in our implementation of SEET, and that in fact the precision achieved with SEET along the entire range from weak to strong correlations is comparable to state-of-the-art quantum chemistry wave function methods, to which we carefully compare. As a testbed, we use small molecular examples, for which exact or nearly exact reference results within a given basis set are available.

The remainder of this paper proceeds as follows. In section~\ref{theory}, we briefly discuss the theoretical setup and the SEET functional. We then follow with benchmarking and in depth explanation of the numerical results in Sec.~\ref{results}. Finally, we present a theoretical discussion supporting our results and conclude in Sec.~\ref{conclusions}.

\section{Methods}~\label{theory}
In this paper, we study small molecular systems.
The specification of the interatomic distance and of a finite set of $N$ gaussian orbitals fully determines the Hamiltonian in second quantized form,
\begin{equation}
H=\sum_{ij}^Nt_{ij}a^{\dagger}_{i}a_{j}+\sum_{ijkl}^Nv_{ijkl}a^{\dagger}_{i}a^{\dagger}_{j}a_{l}a_{k}, \label{eqn:realistic_ham1}
\end{equation}
where the operators $a_i^\dagger$ ($a_i$) create (destroy) an electron in orbital $i$, $t_{ij}$ denotes the single-particle contribution, and $v_{ijkl}$ the Coulomb matrix element.

We express physical properties such as energies and single-particle response functions in a statistical mechanics approach based on an approximation to the grand partition function $Z=\text{Tr}e^{-\beta(H-\mu n)}.$ Here, $\beta$ denotes the inverse temperature, $\mu$ the chemical potential, and $n$ the density operator. Our temperatures are chosen low enough that the system has converged to the ground state, and the chemical potential $\mu$ is adjusted to yield the correct particle density. 

Within this framework, a functional $\Phi[G]$ of the Green's function $G$, which contains all linked closed skeleton diagrams,\cite{Luttinger60} is used to express the grand potential as
\begin{align}
\Omega = \Phi - \text{Tr}( \log G^{-1}) - \text{Tr} (\Sigma G),
\end{align}
where the self-energy $\Sigma$ is defined with respect to a non-interacting Green's function $G_0$ via the Dyson equation
\begin{align}
G=G_0 + G_0 \Sigma G.\label{eq:dyson}
\end{align}
The functional formalism is useful because approximations to $\Phi$ that can be formulated as a subset of the terms of the exact $\Phi$ functional can be shown to respect the conservation laws of electron number, energy, momentum, and angular momentum by construction.\cite{Baym61,Baym62} In addition, $\Phi$-derivability ensures that quantities obtained by thermodynamic or coupling constant integration from non-interacting limits are consistent.\cite{Baym62} Functional theory therefore provides a convenient framework for constructing diagrammatic approximations in situations where a direct solution of the Hamiltonian defined in Eq.~\ref{eqn:realistic_ham1} is not possible.

In order to discuss the formalism used in this paper, we first introduce GF2 and GW, then discuss SEET, and finally explain in detail the SEET+GW formalism.

\subsection{GF2}
The self-consistent second order perturbation theory, GF2, is a $\Phi$-derivable diagrammatic approximation that includes all terms up to second order in the interaction.\cite{Phillips14,Phillips15,Kananenka15,Kananenka16,Rusakov16,Welden16,Dahlen05,Dahlen04,dahlenmp2} In addition to the frequency independent Hartree-Fock terms, $\Sigma_{\infty}$, the second order self-energy contains
\begin{align}
\begin{split}
\Sigma^{(2)}_{ij}(\tau)=-\underset{klmnpq}{\sum}G_{kl}(\tau)G_{mn}(\tau)G_{pq}(-\tau)\\
\times\textnormal{v}_{imqk}\bigr(2\textnormal{v}_{lpnj}-\textnormal{v}_{lpjn}\bigr)~,
\end{split}
\label{eq:Sigma}
\end{align}
where $G(\tau)$ denotes the fully interacting Green's function in imaginary time that is obtained self-consistently from $\Sigma=\Sigma_\infty+\Sigma^{(2)}$ and the non-interacting Green's function $G_0$ using Eq.~\ref{eq:dyson}. The GF2 approach is performed iteratively, starting from a Hartree-Fock Green's function, until $\Sigma$ or the total electronic energy is converged within a predefined tolerance. 
In contrast to the iterative perturbation theory often used in dynamical mean field theory,\cite{Kajueter96} which for the Hubbard lattice interpolates between perturbation theory and the exact solution in the limit of the separated Hubbard atoms, GF2 only relies on perturbation theory and does not recover the correct dissociated limit.
Precise calculations of the second-order results require several numerical optimizations. Our implementation makes use of adaptive grids for both imaginary time~\cite{Kananenka15} and imaginary frequency~\cite{Kananenka16} Green's functions.

\subsection{GW}
The self-consistent GW method\cite{Hedin65,Stan06} is a diagrammatic approximation formulated in terms of renormalized propagators $G$
and renormalized (``screened'') interactions $W$. Similar to GF2, it can be written as an approximation to the Luttinger-Ward (LW) functional $\Phi$\cite{Luttinger60,Almbladh99} and is therefore thermodynamically consistent and conserving.\cite{Baym62,Almbladh99} However, it does not respect the crossing symmetries.\cite{simons_benchmark2}
The method requires the self-consistent determination of propagators $G$, polarizations $P=GG$, self-energies $\Sigma=-GW$, and screened interactions $W$. The expressions for $G$ and $W$ are determined by the Dyson equations
\begin{align}
G= G_0-G_0\Sigma G \;, \qquad \qquad  W =v+ vPW \, ,
\label{Dyson}
\end{align}
where $G_0$ and $v$ are the bare electronic propagator and Coulomb interactions defined in Eq.~\ref{eqn:realistic_ham1}.

Self-consistent GW is only exact to first order in $v$, as the second-order exchange diagram is neglected. 
Our implementation of this approximation closely follows Refs.~\onlinecite{Almbladh99,Stan06,Stan09,Phillips14,simons_benchmark2}.
The Green's function is initialized using the Hartree-Fock result. We then construct the polarization $P=GG$ and obtain $W$ from Eq.~(\ref{Dyson}).
After computing the GW self-energy $\Sigma^{GW}=-GW$, we obtain the updated $G$ by solving Dyson's equation, thus closing the self-consistency loop. 

In order to reduce the size of $W$, we perform a Cholesky decomposition~\cite{Koch03,simons_benchmark2} and 
truncation on $v$. This vastly reduces the numerical effort (for related decompositions see Ref.~\onlinecite{Caruso13}). Adaptive imaginary time~\cite{Kananenka15} and frequency grids~\cite{Kananenka16} are essential to accurately represent the GW Green's function and self-energy. 

\subsection{SEET}
The self-energy embedding theory\cite{Zgid15,Tran15,Tran16,Tran17,zgid_njp17} is a conserving approximation to the Luttinger-Ward functional $\Phi$ designed to treat strongly correlated degrees of freedom. It consists of a two-step hierarchy in which the functional of the system is approximated by a solution of the entire system with a weak coupling method, which is then improved by the non-perturbative solution of correlated subsets of orbitals.
In the most simple case, the strongly correlated subsets are disjoint (non-intersecting) and the SEET functional is defined as
\begin{align}\label{eq:SeetPhi}
\Phi^\text{SPLIT}_\text{SEET} = \Phi^\text{tot}_\text{weak} + \sum_{i=1}^{M} \Big(\Phi_\text{strong}^{A_i}-\Phi_\text{weak}^{A_i}\Big).
\end{align}
Here $\Phi^\text{tot}_\text{weak}$ denotes the approximation of the $\Phi$ functional of the entire system using a weak coupling technique. The index $i$ enumerates the $M$ non-intersecting,  correlated subsets of orbitals $A_i$, and $\Phi^{A_i}$ is the functional evaluated within the orbital subset $A_i$, using a weak coupling or a non-perturbative method.

It is much more general to consider multiple intersecting orbital subspaces. In this case, as described in Ref.~\onlinecite{Tran17}, the SEET functional generalizes to
\begin{align}\label{eq:seet_mix_func} 
\Phi^\text{MIX}_\text{SEET}=\Phi^\text{tot}_\text{weak}
&+\sum_{i}^M(\Phi^{A_i}_\text{strong}-\Phi^{A_i}_\text{weak})\\
&-\sum_{i<j}^M(\Phi^{A_i\cap A_j}_\text{strong} -\Phi^{A_i\cap A_j}_\text{weak})   \nonumber \\ \nonumber
&+\sum_{i<j<k}^M(\Phi^{A_i\cap A_j\cap A_k}_\text{strong}-\Phi^{A_i\cap A_j\cap A_k}_\text{weak})-\cdots,
\end{align}
where the additional terms account for the double counting of overlapping strongly correlated subspaces. The ellipse denotes summations over intersections of increasing numbers of subspaces, which enter with alternating signs.\cite{Tran17}
The indices $i,j,k,\dots=1,\dots,M$ are used to enumerate intersecting orbital subspaces.

The solution of the SEET equations proceeds according to Refs.~\onlinecite{Tran16} and~\onlinecite{zgid_njp17}. After an initial self-energy of the system is obtained within the weak coupling method and correlated subspaces are identified, the algorithm alternates between solving impurity problems with hybridization functions determined by the weak coupling part, and solving the entire systems with propagators that contain impurity self-energies. All impurity problems can be solved in parallel, making this formalism ideally suited for high-performance computing environments.

In SEET, unlike in  DMFT-type approximations, the strongly correlated orbitals are not determined a priori, and are typically not localized. Rather, they are determined after the solution of the system using the weak coupling method. We found that diagonalizing the one-body density matrix allows us to identify orbitals that have partial occupations significantly different from 0 or 2. These partial occupations  can be used to guide the selection of strongly correlated orbitals. Since the selection is done based on occupation numbers obtained from weakly correlated method, it can happen that many orbitals can have similar occupations. Then visualizing the orbitals and analyzing the contributions of parent atomic orbitals to each of the partially occupied natural orbitals can help with the final selection.
By gradually adding more of the strongly correlated orbitals to the correlated subspace, convergence to the exact limit can be observed.\cite{Zgid15,Tran16,Tran17}

In the following, we compare SEET results based on the two weak coupling methods, GW and GF2. For the impurity problems containing strongly correlated orbitals, we use a full configuration interaction (CI) impurity solver and its truncated versions.\cite{Zgid11,Zgid12} The functional $\Phi^\text{tot}_\text{weak}$  then becomes $\Phi^\text{tot}_\text{GF2}$ and $\Phi^\text{tot}_\text{GW}$, respectively, and the functional $\Phi^{A_i}_\text{strong}$ is $\Phi^{A_i}_\text{CI}$.

Several self-consistent iterative procedures developed in other contexts have been shown to suffer from problems such as convergence to an unphysical fixed point\cite{Kozik15} or causality violations.\cite{Haule17_h2} While we cannot exclude that such problems could occur in principle, we have not observed any convergence to unphysical solutions or non-causal Green's functions or self-energies.

\subsection{SEET with GW}
The general algorithmic structure of SEET is described in Refs.~\onlinecite{Zgid15,Tran16}. 
Here, we highlight some aspects of SEET in combination with GW.
\subsubsection{Orbital choice}\label{sec:transformation}
The self-energy embedding can be performed in any orthogonal basis, including in a localized basis [symmetrized atomic orbitals (SAOs) or localized molecular orbitals (LMOs)] or in an energy basis [natural orbitals (NOs) or molecular orbitals (MOs)]. Where we use NOs obtained from the one-body density matrix of the self-consistent GW or GF2 procedure, we select those orbitals to the strongly correlated space which have occupation numbers significantly different from 0 or 2. This space can then be split into multiple intersecting or non-intersecting smaller orbital subspaces with fewer correlated orbitals.
After the selection of strongly correlated orbitals is done, quantities such as $\Sigma^\text{tot}_\text{weak}$,  $[\Sigma^\text{tot}_\text{weak}]^{A_i}$, $[G^\text{tot}_\text{weak}]^{A_i}$, and $[t]^{A_s}$ are expressed in the new basis. 
Here $[X^\text{tot}_\text{weak}]^{A_i}$ denotes the quantity $X$ first obtained for the whole system (denoted with superscript ``tot'') and then restricted to the orbital subspace $A_i$.
Only a subset of the transformed Coulomb integrals $[\tilde{v}]^{A_s}$, namely those where all four orbital indices belong to the subset $A_i$, needs to be evaluated. Here, $v$ stands for all two-body Coulomb integrals in the AO basis while $\tilde{v}$ stands for all the two-body Coulomb integrals in the new basis.

\subsubsection{Double counting}
In SEET, the subtraction of doubly counted diagrams is rigorously defined and unique. No double counting `problem' as in {\it e.g.} LDA+DMFT exists.\cite{Karolak10} We subtract GF2 or GW self-energy evaluated exclusively within each orbital subset $A_s$. In case of GF2, the double counted self-energy is evaluated for $i,j\in A_s$ as,
%\begin{widetext}
\begin{align}\label{eq:Sigma_gf2_dc}
[\Sigma^{DC}_\text{GF2}(\tau)]^{A_s}_{ij}&=-\!\!\!\!\!\underset{klmnpq \in A_{s}}\sum [G^\text{tot}_\text{GF2}(\tau)]^{A_s}_{kl} [G^\text{tot}_\text{GF2}(\tau)]^{A_s}_{mn} \\ \nonumber \times&[G^\text{tot}_\text{GF2}(-\tau)]^{A_s}_{pq} 
 [\tilde{v}]^{A_s}_{imqk} \bigr(2[\tilde{v}]^{A_s}_{lpnj}-[\tilde{v}]^{A_s}_{lpjn}\bigr)~.
\end{align}
%\end{widetext}
The Coulomb integrals used in this evaluation are the transformed integrals $[\tilde{v}]^{A_s}$ belonging to the orbital subset $A_s$. The Green's functions, similarly, are the transformed Green's functions $[G^\text{tot}_\text{GF2}(\tau)]^{A_s}_{kl}$ belonging to the orbital subset $A_s$. 

A similar procedure is used in GW. Here, the decomposed integrals $v_{ijkl}=\sum_{\mu}\bar{v}_{ij}^{\mu}\bar{v}_{kl}^{\mu}$ are transformed to the basis of choice (most frequently natural orbitals) in each orbital subset $A_s$, yielding $[\tilde{v}]^{A_s}_{mnpq}=\sum_{\mu}\bar{\tilde{v}}_{mn}^{\mu}\bar{\tilde{v}}_{pq}^{\mu}$, where $m,n,p,q$ orbital indices belong to orbital subset $A_s$. 
The polarization is then evaluated in the orbital subspace $A_s$ using these transformed integrals, 
\begin{equation}
\Pi(\tau)_{\mu \nu}=- 2\underset{ijlm \in A_{s}}\sum \bar{\tilde{v}}_{il}^{\mu} \bar{\tilde{v}}_{jm}^{\nu}[G^\text{tot}_\text{GW}(\tau)]^{A_s}_{ij}[G^\text{tot}_\text{GW}(-\tau)]^{A_s}_{lm}.
\end{equation}

Subsequently, this polarization diagram is used to evaluate $W(i\omega)=[1-\Pi(i\omega)]^{-1}$ and, after its Fourier transform, the GW self-energy exclusively in the subset $A_s$ is evaluated as
\begin{align}\label{eq:GW_dc}
[\Sigma^{DC}_\text{GW}(\tau)]^{A_s}_{ij}=-\underset{lm \in A_{s}}\sum \sum_{\mu \nu} \bar{\tilde{v}}_{il}^{\mu} \bar{\tilde{v}}_{jm}^{\nu}[G^\text{tot}_\text{GW}(\tau)]^{A_s}_{lm}[W(\tau)]_{\mu \nu}.
\end{align}
While the Green's functions, $[G^\text{tot}_\text{GW}(\tau)]^{A_s}_{lm}$, necessary for the evaluation of the GW self-energy are constructed as a truncation of the total Green's function to the subset $A_s$, the polarization diagram $\Pi(\tau)$ and $W(\tau)$ necessary to construct the double counting corrections are evaluated using the truncated Green's functions and transformed integrals for the subset $A_s$. We emphasize that they are not simply truncations of $\Pi^\text{tot}_\text{GW}(\tau)$ or $W^\text{tot}_\text{GW}(\tau)$ to the subset $A_s$.  Such a definition of these quantities would be incorrect and would likely result in a causality violation of the total self-energy.

\subsubsection{Total self-energy}
The total self-energy (containing both frequency dependent and independent parts $\Sigma=\Sigma_\infty+\Sigma(i\omega)$) of strongly correlated orbitals in each of the subsets $A_s$ has the form
\begin{align}\label{eqn:sigma_tot}
[\Sigma]^{A_s}_{ij}=[\Sigma^\text{tot}_\text{weak}]^{A_s}_{ij}+([\Sigma_\text{strong}]^{A_s}_{ij}-[\Sigma^{DC}_\text{weak}]^{A_s}_{ij}),
\end{align}
where the method describing ``weak'' correlation is either GF2 or GW. The double counting correction is subtracted in the term $[\Sigma^{DC}_\text{weak}]^{A_s}$, which is given by Eq.~\ref{eq:Sigma_gf2_dc} in the case of GF2 and by Eq.~\ref{eq:GW_dc} in case of GW. The self-energy $[\Sigma_\text{strong}]^{A_s}$ is obtained by solving a quantum impurity problem. The construction of the impurity model for SEET will be discussed in Sec.~\ref{sec:AIM_SEET}. 

We can rewrite Eq.~\ref{eqn:sigma_tot} as
\begin{align}\label{eq:sigma_embedding}
[\Sigma]^{A_s}_{ij}=[\Sigma^\text{embedding}_\text{weak}]_{ij}+[\Sigma_\text{strong}]^{A_s}_{ij}.
\end{align}
Here $[\Sigma^\text{embedding}_\text{weak}]_{ij}$ provides an effective correction to $[\Sigma_\text{strong}]^{A_s}_{ij}$. While $[\Sigma_\text{strong}]^{A_s}_{ij}$ has all contributions from the subset of interactions $[\tilde{v}]^{A_s}_{ijkl}$ where $i,j,k,l \in A_s$, the correction $[\Sigma^\text{embedding}_\text{weak}]_{ij\in A_s}$ accounts for  all those diagrammatic contributions obtained using $\tilde{v}_{mnkl}$ where at least one of the indices is outside the orbital subset $A_s$ but at the same time at least one of the indices belongs to $A_s$.

\subsubsection{Impurity problem in SEET}\label{sec:AIM_SEET}
In SEET, the Green's function restricted to subspace $A_s$ is written as
\begin{equation}\label{eq:subsetprop}
[G^\text{tot}]^{A_s} = \Big( [G_0^{-1}]^{A_s} - [\Sigma]^{A_s} - \Delta \Big)^{-1}, 
\end{equation}
where $\Delta$ describes a hybridization between the orbital subset $A_s$ and the environment. The self-energy $[\Sigma]^{A_s}$ is defined by Eq.~\ref{eqn:sigma_tot} and $[G_0^{-1}]^{A_s}$ is the inverse of the bare Green's function in the orbital subspace $A_s$, defined as
\begin{equation}\label{eq:g0eq}
[G_0^{-1}]^{A_s}=(i\omega+\mu){\bf 1}-[t]^{A_s}.
\end{equation}
Here $[t]^{A_s}$ denotes a kinetic energy operator truncated to the subspace $A_s$.

The embedding of correlated orbitals into a background of other orbitals requires the solution of a quantum impurity problem with an impurity solver algorithm, which provides the interacting self-energy.
The impurity $\mathcal{G}_0$ Green's function is defined as 
\begin{align}\label{eqn:g0}
\mathcal{G}_0^{-1}= [G_0^{-1}]^{A_s}-[\Sigma^{NDC}_\infty]^{A_s} - \Delta,
\end{align}
where the frequency independent part of the self-energy is constructed as
\begin{align}
[\Sigma^{NDC}_\infty]^{A_s}&=[\Sigma^{\text{tot}}_{\infty}]^{A_s}-[\Sigma^{DC}_\infty]^{A_s} \\
[\Sigma^{\text{tot}}_{\infty}]^{A_s}_{ij \in A_s} & =\sum_{kl} \gamma_{kl}(\tilde{v}_{ijkl}-0.5 \tilde{v}_{ilkj}) \\
[\Sigma^{DC}_{\infty}]^{A_s}_{ij \in A_s}& =\sum_{kl \in A_s}\gamma_{kl}(\tilde{v}_{ijkl}-0.5 \tilde{v}_{ilkj}),
\end{align}
with the density matrix $\gamma$ obtained from GF2 or GW. Note that $[\Sigma^{DC}_{\infty}]^{A_s}$ is used to subtract double counting coming from the subspace $A_s$.

A quantum impurity solver will obtain an expression for a correlated $[G^\text{imp}]^{A_s}$ given $\Delta$, $[\Sigma^{NDC}_\infty]^{A_s}$,  and $G_0$ (Eq.~\ref{eq:g0eq}) as well as a subset of interactions $[\tilde{v}]^{A_{s}}$. As we formulate SEET in this work for the Luttinger Ward $\Phi$ functional, rather than for the Almbladh $\Psi$ functional, the impurity interactions remain instantaneous.

Using the impurity problem Dyson equation, the self-energy for a strongly correlated orbital subset is obtained as  
\begin{equation}\label{eqn:dyson_imp}
[\Sigma_\text{strong}]^{A_s}=\mathcal{G}_0^{-1}-([G^\text{imp}]^{A_s})^{-1}.
\end{equation}
Once this strongly correlated $[\Sigma_\text{strong}]^{A_s}$ is known, the total self-energy, $[\Sigma]^{A_s}$, in subspace $s$ is evaluated according to Eq.~\ref{eqn:sigma_tot}.

In general, for a realistic system described by a Hamiltonian with non-local one- and two-body interactions, the hybridization $\Delta$ in the orbital subspace $A_s$ contains off-diagonal terms ({\it i.e.} $\Delta_{ij}\neq 0$ for $i\neq j$).
However, let us note that hybridizations in the MO or NO basis are often almost diagonal, so that these bases either completely eliminate the off-diagonal terms or greatly reduce them. Additionally, the magnitude of the diagonal hybridization elements is often substantially smaller than those obtained in the local basis of atomic orbitals. We will come back to these points in Sec.~\ref{sec:h2_hyb}.

One of the few impurity solvers able to treat general hybridization functions and general interactions at low temperature is the configuration-interaction (CI) method or its truncated versions (RASCI)~\cite{Zgid11,Zgid12}. (For later implementations of these methods see Refs.~\onlinecite{Lin13,Haverkort14,Go17}).  When applied to impurity problems the CI method relies on a fit of the hybridization function to a discretized set of non-interacting bath levels, which we perform on the imaginary axis.

\subsection{SEET self-consistency}
The main algorithmic step in the SEET self-consistency are as follows (see appendix of Ref.~\onlinecite{zgid_njp17} as well as Ref.~\onlinecite{Tran16} for a more extensive description).
\begin{enumerate}
\item Perform a Hartree-Fock (HF) or density functional theory (DFT) calculation. Then, using this starting point, perform a self-consistent GF2 or GW calculations yielding $G_\text{GF2}^\text{tot}$ or $G_\text{GW}^\text{tot}$ for the entire system.  
\item \label{dmeval_iter} If using a natural orbital basis, evaluate  and diagonalize the GF2 or GW density matrix to obtain natural orbitals and occupation numbers. We define strongly correlated orbitals as those with occupations numbers significantly different from 0 or 2. If using a local basis (SAO), choose the correlated orbitals based on spatial criteria.
\item Transform all Green's function and self-energies as well as $t$ and $v$ integrals~\footnote{only a subset of $v$ integrals has to be transformed to the new orthogonal basis.} from the atomic orbital (AO) basis to an orthogonal orbital basis such as, MOs, NOs, LMOs, and SAOs. For details see Sec~\ref{sec:transformation}.
\item Construct using Eq.~\ref{eq:Sigma_gf2_dc} or Eq.~\ref{eq:GW_dc} the GF2 or GW self-energy which will be subtracted  to eliminate the double counting.
\item Construct the impurity Hamiltonian, in which two-electron term is a subset of bare Coulomb interaction. Note that this interaction matrix was already transformed to the new orthogonal basis. For details see Sec.~\ref{sec:AIM_SEET}.
\item Perform the inner self-consistency loop:
  \begin{enumerate}
  \item For every orbital subset $A_s$, use an impurity solver (RASCI/FCI) to obtain the impurity Green's function $[G^\text{imp}]^{A_s}$ and extract the impurity self-energy $[\Sigma_\text{strong}]^{A_s}$. At the first iteration the hybridization $\Delta$ is initialized using GF2 or GW quantities. 
  \item Set up the total self-energy according to Eq.~\ref{eqn:sigma_tot}. 
  \item Reconstruct the total Green's function via the Dyson equation and adjust the chemical potential to obtain a correct electron number for the  whole system.
  \item Update the hybridization $\Delta$ using the new total Green's function and self-energy.
  \item Go back to step 6(a) and iterate until convergence is reached.
  \end{enumerate}
\item \label{outloop1} Pass the converged Green's function to a GF2 or GW calculation and perform only a single iteration.
\item \label{outloop2} Go back to step~\ref{dmeval_iter} and iterate until outer loop convergence is reached.
\end{enumerate}
In practice, we found that closing the outer self-consistency loop and performing steps \ref{outloop1} and \ref{outloop2} has little effect when a natural orbital basis is used.  In the SAO basis, the outer loop has a larger effect and is necessary~\cite{Tran16}.

\section{Results}\label{results}
Unless otherwise noted, the \textsc{ORCA} program \cite{orca} was used for all calculations using standard methods such as second-order M{\o}ller-Plesset perturbation theory (MP2), FCI, complete active space configuration interaction (CASCI), and $n$-electron valence state second-order perturbation theory (NEVPT2) \cite{Angeli01,Angeli02}.
A locally modified version of the \textsc{DALTON} code \cite{dalton} was employed to generate a restricted HF reference prior to GF2 and GW calculations.

For all hydrogen chains examined, GF2 and GW electronic energies were converged to $10^{-5}$ Ha. All SEET(CI/GF2) and SEET(CI/GW) calculations were then executed with the convergence threshold of $10^{-4}$ Ha. 

The inverse temperature $\beta$ was set at 100 Ha$^{-1}$ or 200 Ha$^{-1}$ depending on the system and geometry. The Matsubara freqency grid was generated using the spline interpolation~\cite{Kananenka16} with the maximum number of points varying between 20,000 and 50,000. 

We use the following shorthand notation, SEET(method strong/method weak)-[$K$o] stands for SEET employing ``method weak'' as a low-level, weak correlation method for the entire system. The impurity problems composed of $K$ strongly correlated orbitals and a number of bath orbitals necessary to fit the hybridization are solved using ``method strong''.
Moreover, the two variants of SEET functional are denoted as SEET(method strong/method weak)-split[$L\times K$o] and SEET(method strong/method weak)-mix[$K$o]. The former employs ``method strong'' to solve impurity problems created from $L$ non-intersecting orbital subspaces containing $K$ strongly correlated orbitals while the latter is executed for intersecting orbital subspaces containing $K$ strongly correlated orbitals. 

\subsection{H$_2$ molecule}

\subsubsection{Energetics}

\begin{figure} [htb]
  \includegraphics[width=7.0cm,height=9cm]{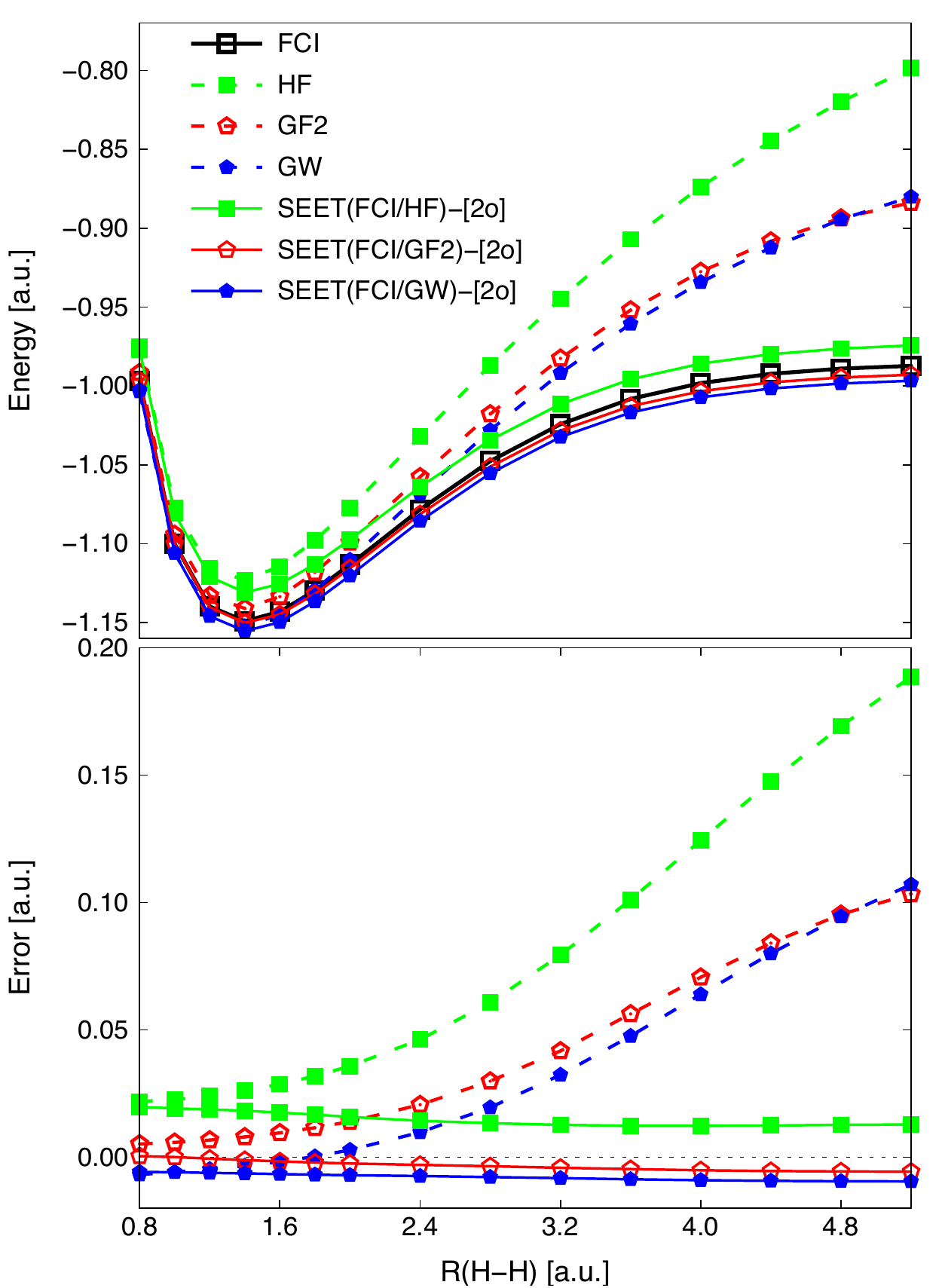} \caption{Upper panel: Potential energy curves of H$_2$ in TZ(Dunning) basis. Lower panel: Errors with respect to the FCI reference.}
  \label{fig:H2TZ}
\end{figure}

We first examine the implementation of SEET with GW as a weak correlation method for the simplest molecular system, H$_2$ in TZ(Dunning) basis~\cite{TZDunning}. In this system, there are 2 electrons in 6 orbitals.
Our results using SEET(FCI/HF)-[2o], SEET(FCI/GF2)-[2o], and SEET(FCI/GW)-[2o]  are summarized in Fig.~\ref{fig:H2TZ}.
The full configuration interacting (FCI) method is used as an exact solution within this basis set. The weakly correlated methods such as restricted HF, GF2 and GW that are based on restricted HF reference  are provided for comparison.

In the weakly correlated limit, around the equilibrium geometry, both GF2 and GW agree qualitatively well with FCI.  In contrast, in the strongly correlated limit, for large interatomic distances, the GF2 and GW are qualitatively incorrect resulting in large errors.
A significant improvement upon all of the weakly correlated methods is achieved
when the SEET(FCI/method weak)-[2o] is executed. For GF2 and GW, the two strongly correlated impurity orbitals are identified as natural orbitals  with partial occupancy at large interatomic distances. These two orbitals can be then systematically followed until short interatomic distances are reached. For HF, we defined strongly correlated orbitals as the highest occupied molecular orbital (HOMO) and the lowest unoccupied orbital (LUMO). 
Though the SEET(FCI/HF) curve is nearly parallel to the FCI curve for large interatomic distances, its errors are still considerable, with the maximum error of 0.019 Ha. These errors arise due to the lack of weak correlation between the strongly correlated impurity orbitals and the remaining (environment) orbitals.
SEET(FCI/GF2) and SEET(FCI/GW), which can capture strong correlations within the impurity orbitals and weak correlations among the environment orbitals and between the environment and strongly correlated orbitals,
yield curves closely following the FCI curve with maximum errors of 0.005 and 0.009 Ha for SEET(FCI/GF2)-[2o] and SEET(FCI/GW)-[2o], respectively.

\begin{figure} [htb]
  \includegraphics[width=7.0cm,height=5.0cm]{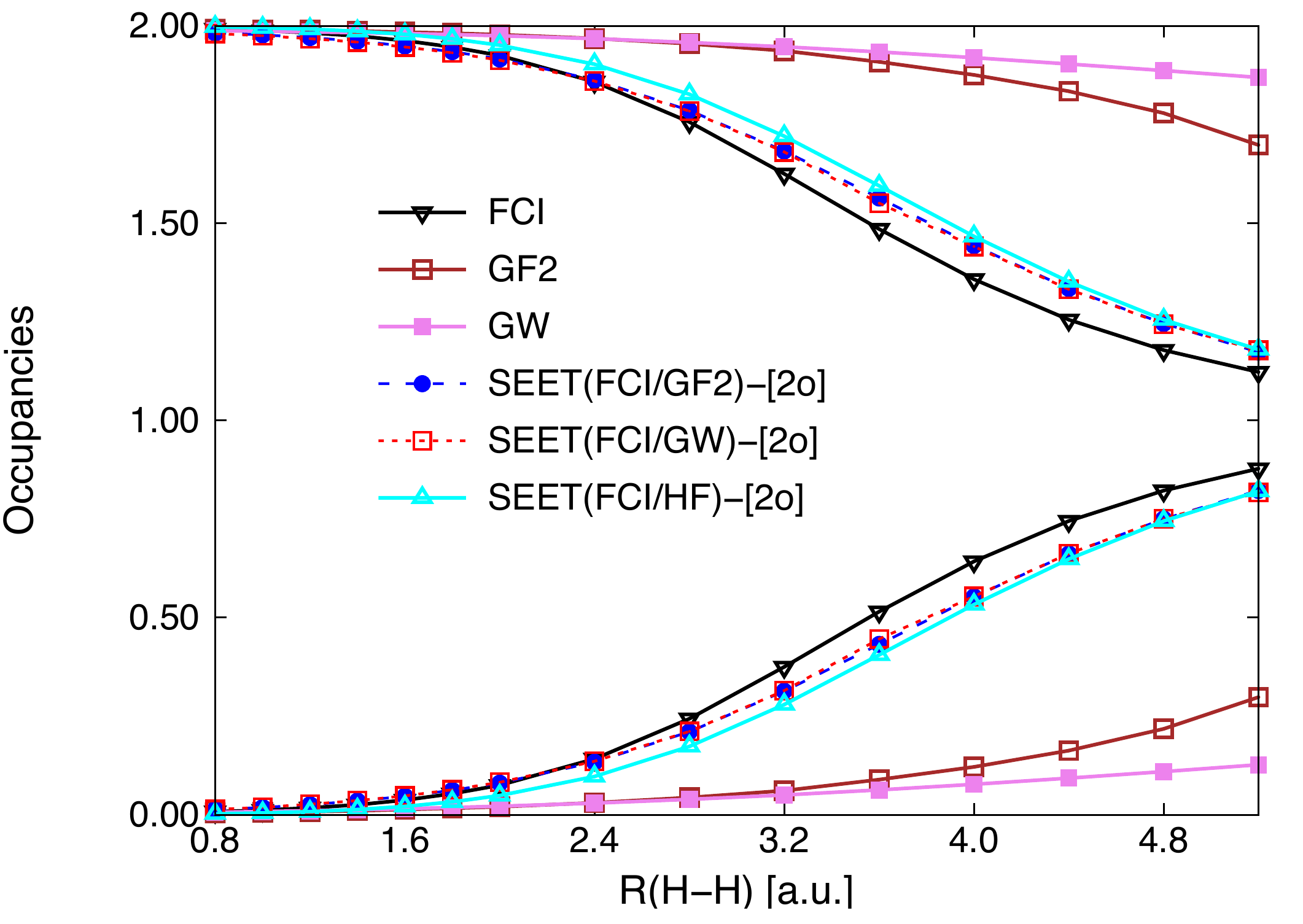}  \caption{Occupation numbers of H$_2$ in TZ(Dunning) as a function of interatomic distance.}
  \label{fig:H2TZ_occ}
\end{figure}

Occupation numbers provide an additional insight into performance or SEET.
In Fig.~\ref{fig:H2TZ_occ}, we plot the occupation numbers as a function of the interatomic distance from low-level methods as well as different variants of SEET. FCI occupation numbers are listed for comparison.
Both GF2 and GW occupation numbers differ from 2 and 0  when the interatomic distance is large but remain significantly different from the FCI reference.
All SEET variants correctly describe the transition from the weakly to strongly correlated regime and SEET occupation numbers follow the trend displayed by FCI very well.

\begin{figure} [htb]
  \includegraphics[width=7.0cm,height=14.0cm]{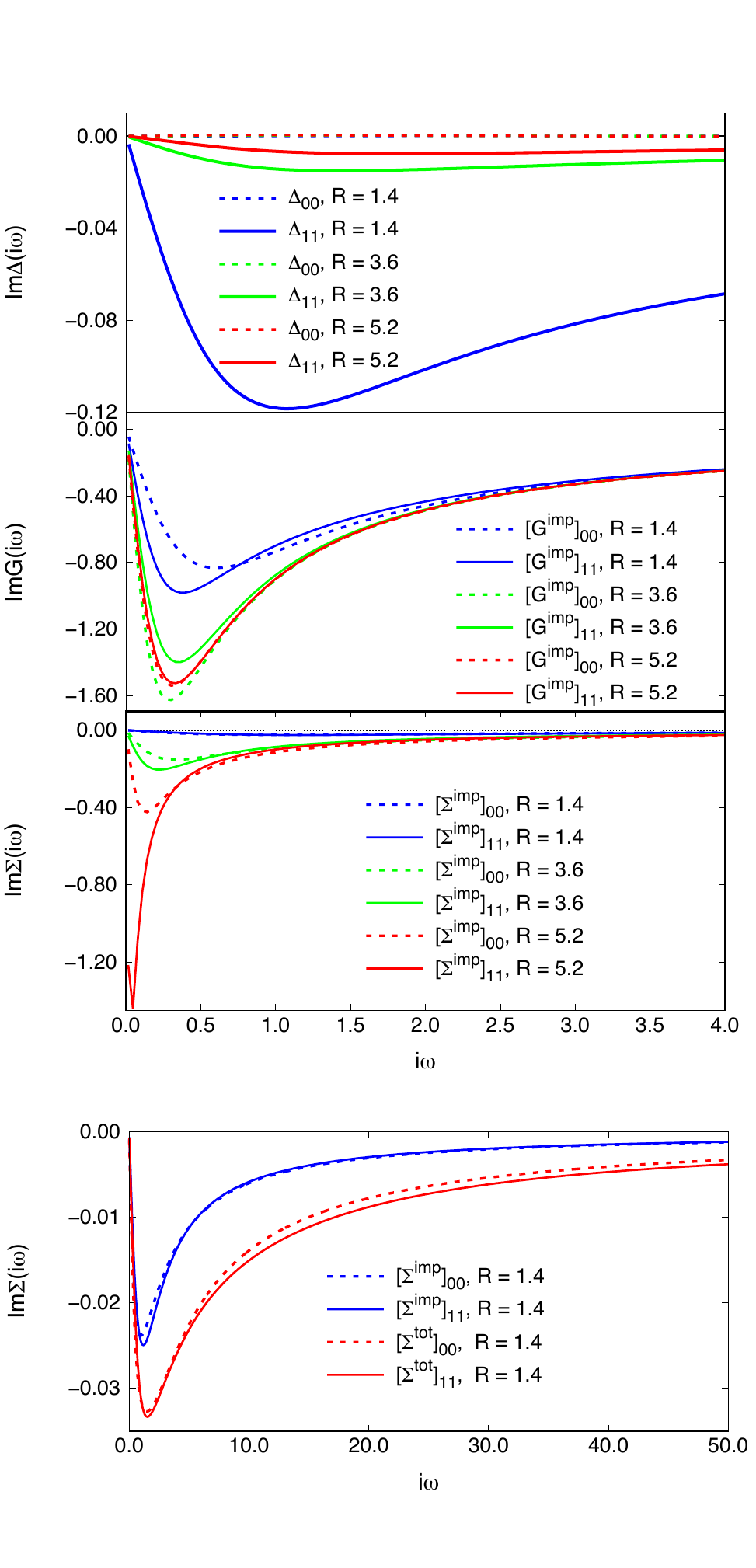} \caption{Upper panels: Imaginary part of the diagonal elements of the hybridization, impurity Green's function, and impurity self-energy  evaluated at various bond lengths for H$_2$ in the TZ(Dunning) basis.  Lower panel: Imaginary part of diagonal elements of the impurity self-energy  and total self-energy (see Eq.~\ref{eq:sigma_embedding}) evaluated at a distance of 1.4 (a.u.) for H$_2$ in the TZ(Dunning) basis. Natural orbitals were used for all calculations.}
  \label{fig:causal}
\end{figure}

\subsubsection{Green's function, self-energy, and hybridization}\label{sec:h2_hyb}

In this section we will discuss both the hybridization and impurity self-energy to provide insight into the SEET performance. 

SEET employs an accurate, non-perturbative method to solve the impurity problem containing strongly correlated orbitals and an approximate method to treat the remaining weakly correlated orbitals. Consequently, our major goal is to find an orbital basis  that maximizes the magnitude of the impurity self-energy and minimizes the magnitude of the hybridization between the impurity orbitals and the environment.  A small magnitude of the hybridization means that there is only a small electron exchange between the impurity and the environment and essentially only a weak entangelment of the strongly correlated impurity orbitals with the environment. The impurity self-energy, which illustrates many-body effects between the strongly correlated orbitals, can be large since an accurate, non-perturbative method is employed to calculate it. 

Natural orbitals, which are eigenvectors of the one-body density matrix, are an orbital basis that fulfills the above criteria. It is evident form the top panel of Fig.~\ref{fig:causal} that when the orbitals are strongly correlated for $R \geq 3.6$ a.u., the imaginary part of hybridization is small while the imaginary part of the impurity self-energy is large. Only for $R = 1.4$ a.u. the hybridization is larger than the self-energy. However, in this limit, the system is weakly correlated and essentially the weakly correlated method is capable of describing it.
From the bottom panel of Fig.~\ref{fig:causal}, we see that in this weakly correlated limit ($R = 1.4$ a.u.), the magnitude of the frequency dependent impurity self-energy is similar to the total self-energy from Eq.~\ref{eqn:sigma_tot}. Consequently, as expected, when hybridizations are large to achieve quantitative accuracy, it is essential for a well behaved quantitative theory to contain $\Sigma^\text{embedding}_\text{weak}$ as described in Eq.~\ref{eq:sigma_embedding}.

Recently, Ref.~\onlinecite{Haule17_h2} reported non-causal hybridizations for the H$_2$ molecule for large interatomic distances when the GW+DMFT method was employed. In SEET, as evident from Fig.~\ref{fig:causal}, we do not observe any non-causal hybridizations, self-energies, or Green's functions. In fact, at large interatomic distances, when natural orbital basis is employed, the hybridization of the two strongly correlated orbitals with the environment is close to zero. At present it is unclear if the causality issue observed in Ref.~\onlinecite{Haule17_h2} is a deficiency of the implementation or a more fundamental theory problem.

\subsection{H$_{10}$ chain}

Here, we will further focus on a detailed comparison between SEET(CI/GF2) and SEET(CI/GW).
There are two aspects in which SEET(CI/GF2) and SEET(CI/GW) can yield different results that we would like to examine:
\begin{enumerate}[label=(\roman*)]
\item the recovery of strong correlations when the strongly correlated orbitals are partitioned between multiple impurity problems. This happens when there are too many strongly correlated orbitals to contain them in a single impurity that is treated by a non-perturbative solver,
\item the description of weak correlations (outside the strongly correlated space) among weakly correlated orbitals.
\end{enumerate}

\begin{figure*} [htb]
\includegraphics[width=16.0cm,height=8.0cm]{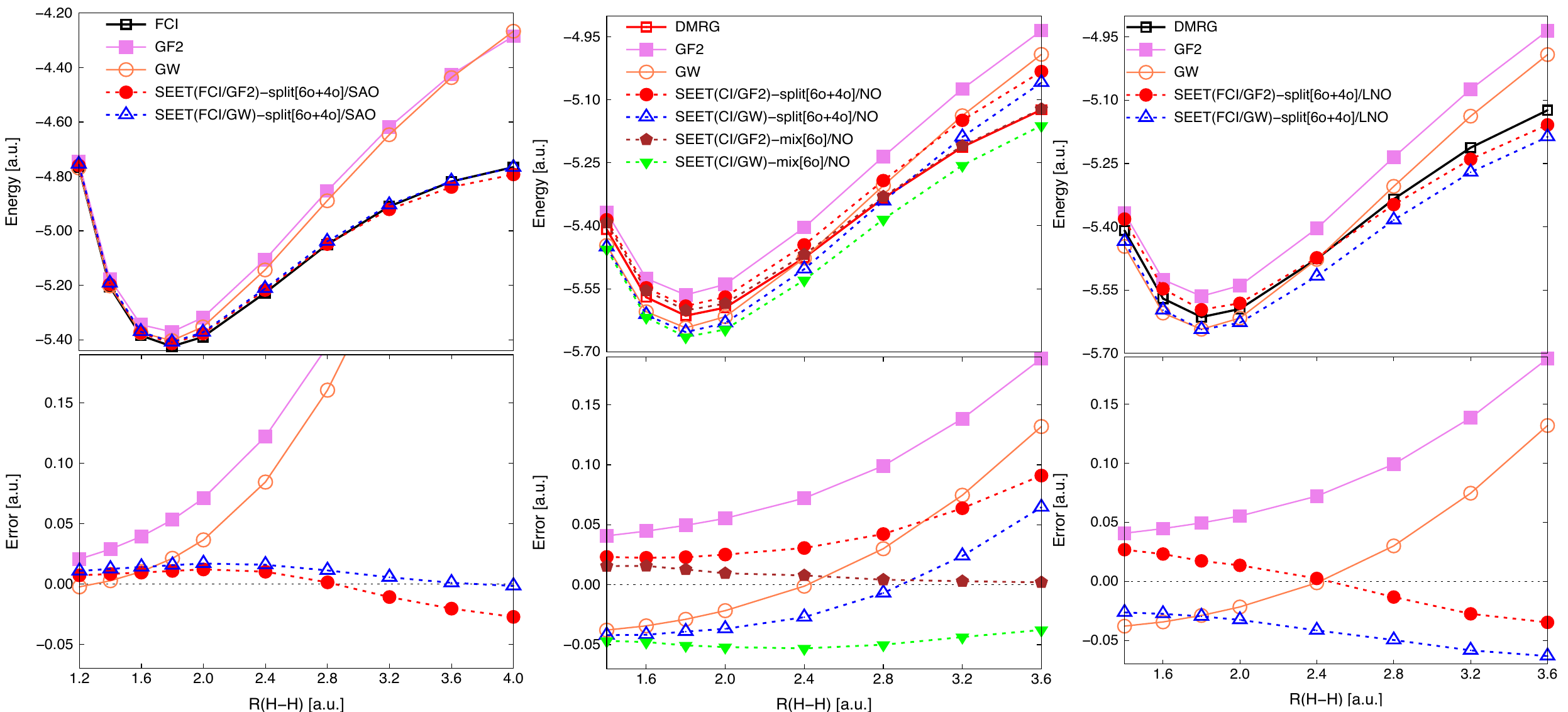} \caption{Left panel: Potential energy curves (upper) and total errors (lower) with respect to the FCI reference for H$_{10}$ chain in STO-6G basis. 
  Middle panel: Potential energy curves (upper) and total errors (lower) with respect to the DMRG reference for H$_{10}$ chain in cc-pVDZ basis.
 SEET calculations were performed  using natural orbital (NO) basis. Right panel: Potential energy curves (upper) and total errors (lower) with respect to the DMRG reference for H$_{10}$ chain in cc-pVDZ basis. SEET calculations were performed using localized natural orbital (LNO) basis. The DMRG reference was taken from Ref.~\onlinecite{simons_benchmark2}.}
\label{fig:H10}
\end{figure*}

To access point (i) we employ a simple H$_{10}$ chain in STO-6G\cite{sto_minimal} basis. In this minimal  basis, for large interatomic distances, all 10 orbitals are strongly correlated. This means that if we use a single impurity to contain all the strongly correlated orbitals, it would need to contain all 10 of them.
Since most solvers, at low temperatures,  cannot handle impurities with more than 5-6 strongly correlated orbitals, we choose to split these 10 orbitals into two smaller orbital subspaces containing 6 and 4 orbitals each.
This means that strong correlations that are potentially arising between these two orbital subspaces are treated only at the pertubative level by GF2 or GW.
The results of such a treatment employing SEET(FCI/GF2)-split[6o+4o]/SAO and SEET(FCI/GW)-split[6o+4o]/SAO are listed in the left panel of Fig.~\ref{fig:H10}.
Here, SAO denotes symmetrically orthogonal atomic orbital basis that results in almost localized orbitals.

As expected, both GF2 and GW capture weak correlation at the quantitative level. However, even though they can partially capture the strong correlations, they are not sufficient to accurately describe the dissociation limit (strongly correlated limit) quantitatively. Both SEET(FCI/GF2)-split[6o+4o] and SEET(FCI/GW)-split[6o+4o]  improve dramatically upon  GF2 and GW results at long distances when strong correlations are present.
 The differences between SEET(FCI/GF2) and SEET(FCI/GW) are more pronounced in the strongly correlated regime than in the weakly correlated one. 
While SEET(FCI/GF2)-split[6o+4o] energies are below the FCI reference past the distance of 2.8 a.u., such energy overestimation is much smaller for SEET(FCI/GW)-split[6o+4o].
Errors at the distance of 4.0 a.u. are --0.027 Ha and --0.002 Ha for SEET(FCI/GF2)-split[6o+4o]  and SEET(FCI/GW)-split[6o+4o], respectively.

To further examine both (i) and (ii) points simultaneously, we employ an H$_{10}$ chain in  cc-pVDZ\cite{dunning_ccpvdz}.
In the  cc-pVDZ basis, there are 50 orbitals in total. Forty  of them are weakly correlated while the remaining ten orbitals, for large internuclear distances, are strongly correlated.
The DMRG reference was taken from Ref.~\onlinecite{simons_benchmark2}.
We start with the examination of  SEET results in NOs which are obtained from the weak correlation method such as GF2 or GW, see the middle panel of Fig.~\ref{fig:H10}. In the NO basis, orbitals remain delocalized. 

In the strongly correlated regime, for large interatomic distances, both GF2 and GW that are based on restricted HF reference largely deviate from the DMRG results and yield underestimated correlation energies. 
In the weakly correlated regime, for short intermolecular distances, the behavior of GW differs from that of GF2. 
In this regime, GF2 potential energy curve is always above the DMRG reference, while the GW curve is below the DMRG reference, yielding overestimated correlation energies.

Performing SEET and splitting the strongly correlated orbitals into non-intersecting orbital subspaces containing 4 and 6 strongly correlated orbitals significantly improves the results.
In particular, SEET(CI/GF2)-split[6o+4o]/NO noticeably reduces GF2 errors for the whole range of distances.
SEET(CI/GW)-split[6o+4o]/NO, however, only improves GW energies at long distances where the strong correlations are important. At short interatomic distances, no visible improvement due to SEET(CI/GW)-split[6o+4o]/NO is observed.
The overestimation of GW correlation energy at short distances cannot be improved by SEET. This is because in this regime, correlations are dominated by the weak correlations between the strongly correlated and weakly correlated orbitals, rather than only the strong correlations within the space of the strongly correlated orbitals.

Now, let us focus on longer interatomic distances for the example of H$_{10}$ chain in  cc-pVDZ\cite{dunning_ccpvdz} basis using NOs.
Here, in the SEET-split scheme strong correlations between both the orbitals subspaces containing 4 and 6 orbitals become very important allowing us to examine point (i) in detail. Simultaneously, in this larger basis there are of course correlations between the strongly correlated orbitals and the remaining weakly correlated ones giving rise to correlations described by point (ii). However, for long interatomic distances, the missing strong correlations appearing in the SEET-split scheme dominate any effects appearing due to the weak correlations. 

For large interatomic distances, both SEET(CI/GF2)-split[6o+4o]/NO  and SEET(CI/GW)-split[6o+4o]/NO are unable to sufficiently capture the strong correlations present between orbital subspaces containing 4 and 6 orbitals.
Both SEET(CI/GF2)-split[6o+4o]/NO and SEET(CI/GW)-split[6o+4o]/NO result in large nonparallelity errors (NPEs)~\footnote{Nonparallelity errors (NPEs) are defined as the difference between the largest and smallest errors with respect to the reference data. Methods with small NPEs may not give an exact total energy but result in many useful data since the shift in total energy is then only a constant shift.}
of 0.068 Ha  and 0.105 Ha, respectively. 
These significant NPEs are caused by the failure of SEET-split scheme in NO basis to recover important  correlations between two non-intersecting strongly orbital subspaces at large interatomic distances.

To investigate if this difficulty can be removed, we 
 employed the generalized version of SEET (SEET-mix) allowing for the presence of  intersecting strongly correlated orbital subspaces~\cite{Tran17}.
NPEs of SEET(CI/GF2)-mix[6o]/NO and SEET(CI/GW)-mix[6o]/NO 
are now 0.013 Ha and 0.015 Ha, respectively. 
These NPEs are one magnitude smaller than those of SEET-split scheme, indicating a proper description of the strong correlations between intersecting orbital subspaces. 

While the NPEs are similarly small for both SEET-mix scheme with GF2 and GW, it is worth noting that the  SEET(CI/GF2)-mix[6o]/NO curve is in a very good agreement with the DMRG reference with the maximum absolute error of 0.016 Ha. 
In contrast, due to the huge overestimation of GW correlation energy, reflected by its very low total energies at short distances, the SEET(CI/GW)-mix[6o]/NO curve remains parallel to however much  below the DMRG reference. 
We also observed a similar behavior in the already considered  H$_2$ example in TZ(Dunning) basis.
Consequently, once the SEET-mix scheme is applied the weak correlations described by point (ii) are responsible for the overall quality of results.

When examining the example of H$_{10}$ chain in the minimal STO-6G basis using local SAOs, we observed that in the strongly correlated regime, SEET(FCI/GW)-split gives better results than SEET(FCI/GF2)-split.
It is therefore interesting to further examine the behavior of  SEET(FCI/GW)-split versus SEET(FCI/GF2)-split when localized orbitals are employed for a larger basis set than the minimal one.
To consider such a case, we continue to focus on the H$_{10}$ chain in cc-pVDZ basis. 
Ten valence orbitals from the set of fifty NOs are localized using the Boys localization procedure~\cite{Boys60}. 
We will call these orbitals the localized natural orbitals (LNOs).
Let us note here that both Boys and Wannier~\cite{} localizations are equivalent when finite systems are considered.

Here, yet again our major goal is to examine points (i) and (ii) simultaneously. First, we illustrate how strong correlations between non-intersecting subspaces are recovered when using LNO basis.
The results are displayed on the right panel of Figure~\ref{fig:H10}. 

While there is no improvement of NPE for SEET(FCI/GF2)-split[6o+4o] (0.062 Ha in LNOs compared to 0.068 Ha in NOs), NPE for SEET(FCI/GW)-split[6o+4o] in LNOs (0.037 Ha) is considerably reduced from its counterpart in the delocalized NO basis (0.105 Ha).

Similarly to the case of the H$_{10}$ chain in STO-6G, the SEET(FCI/GF2)-split[6o+4o]/LNO curve is below the DMRG one at long distances implying an overcorrelation in SEET with GF2.
While the NPE of SEET(FCI/GW)-split[6o+4o]/LNO is small, the total energy is overestimated for all intermolecular distances.

In all cases considered, irrespective of  the orbital basis or basis set, we have observed a systematic overcorrelation of total energies when any variant of SEET(CI/GW) is employed. 
The overcorrelation present in GW becomes more pronounced when a larger basis set is used. 
Consequently, in the SEET(CI/GW) approach, the overcorrelation present in GW itself is similarly affecting the description of weakly correlated orbitals outside the strongly correlated orbital space (point (ii)). SEET(CI/GW)-mix scheme was not able to remove these overcorrelations.

For SEET(CI/GF2)-split, we observed the total energy overcorrelation only at large interatomic distances when the localized basis SAO or LNO was applied. In the NO basis, we observed underestimation of correlation energies. These problems in both localized (SAO, LNO) or energy (NO) basis can be partially relieved when SEET(CI/GF2)-mix scheme is employed~\cite{Tran17}. 

\subsection{H$_{50}$ chain}

\begin{figure} [htb]
  \includegraphics[width=6.5cm,height=9.0cm]{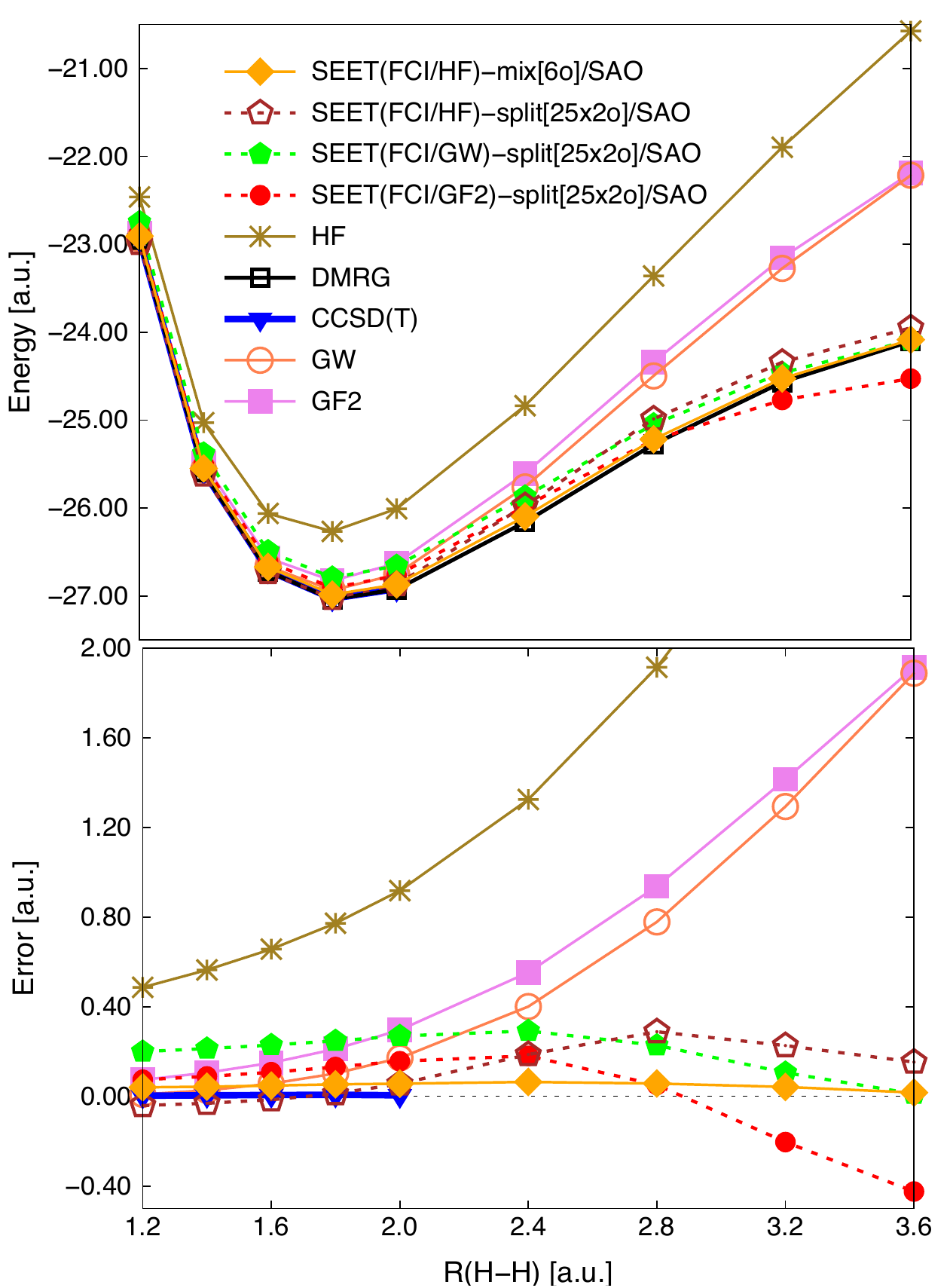} \caption{Upper panel: Potential energy curves of H$_{50}$ chain in STO-6G basis. Lower panel: Errors from the DMRG reference. DMRG data were taken from Ref.~\onlinecite{Hachmann:jcp2006-h50dmrg}.}
  \label{fig:H50}
\end{figure}

We now apply the SEET(CI/GW) implementation to a much longer hydrogen chain, H$_{50}$ in the STO-6G basis. In quantum chemistry, this is a well-known benchmark for strongly correlated methods since for large interatomic distances the full strongly correlated space is large and contains 50 electrons in 50 orbitals. The reference solution is available from DMRG calculations.\cite{Hachmann:jcp2006-h50dmrg} For this system, traditional methods suitable for weakly correlated systems, such as CCSD(T), are unable to converge past the distance of 2.0 a.u.\cite{Hachmann:jcp2006-h50dmrg}
We employ here the SAO basis, which yields almost localized orbitals. 
In SEET-split calculations, the full strongly correlated space of 50 orbitals is split into 25 non-intersecting subspaces composed of 2 orbitals each.
Correlations arising between these non-intersecting subspaces are then treated either by GF2 or GW.
The resulting potential energy curves and total errors with respect to the DMRG reference are summarized in Fig.~\ref{fig:H50}. 

In the weakly correlated regime ($R < 2.0 $ a.u.), GF2 and GW give results of a similar quality to SEET(FCI/GF2)-split[25$\times$2o] and SEET(FCI/GW)-split[25$\times$2o]. Both SEET(FCI/HF)-split[25$\times$2o] and SEET(FCI/HF)-mix[6o] significantly improve the HF result.
In the strongly correlated regime ($R \geq 2.0 $ a.u.), both GF2 and GW based on the RHF reference are unable to properly describe strong correlations and display large underestimation of correlation energies.

For these longer distances, SEET(FCI/HF)-split[25$\times$2o], SEET(FCI/GF2)-split[25$\times$2o], and SEET(FCI/GW)-split[25$\times$2o] show a large improvement over the HF, GF2, and GW results, providing energies that are close to the DMRG reference. 
As opposed to SEET(FCI/GF2)-split[25$\times$2o], SEET(FCI/GW)-split[25$\times$2o] has a much smaller overcorrelation in the strongly correlated regime. 
The total error (--0.522 Ha) present in SEET(FCI/GF2)-split[25$\times$2o] at the distance of 4.2 a.u. is one order of magnitude larger than that (--0.055 Ha) obtained with SEET(FCI/GW)-split[25$\times$2o].

At large distances, in the minimal basis set when SAOs are used, these errors arise due to overcorrelations present in SEET-split scheme with both GF2 and GW employed as weak correlation methods that are recovering correlations between non-intersecting subspaces of strongly correlated electrons. At least in the minimal basis, when SAOs are used, this overcorrelation can be removed when SEET(FCI/HF)-split[25$\times$2o] or SEET(FCI/HF)-mix[6o] are performed, for details see Ref.~\onlinecite{Tran17}.
For SEET(FCI/HF)-mix[6o]/SAO that uses intersecting strongly correlated orbital subspaces, the strong correlation between orbitals is recovered well.

Results for SEET(FCI/HF)-split[25$\times$2o] and SEET(FCI/HF)-mix[6o] of this quality are only possible in the minimal basis and when localized orbitals are employed. In a larger basis, where both weakly and strongly correlated orbitals are present and where there are  fewer strongly than weakly correlated orbitals, SEET((FCI/GF2) or SEET((FCI/GW) are necessary to get quantitative results. We will discuss such a case in the next section.

\subsection{N$_2$ molecule}\label{N2result}
\begin{figure} [htb]
  \includegraphics[width=7.0cm,height=5.0cm]{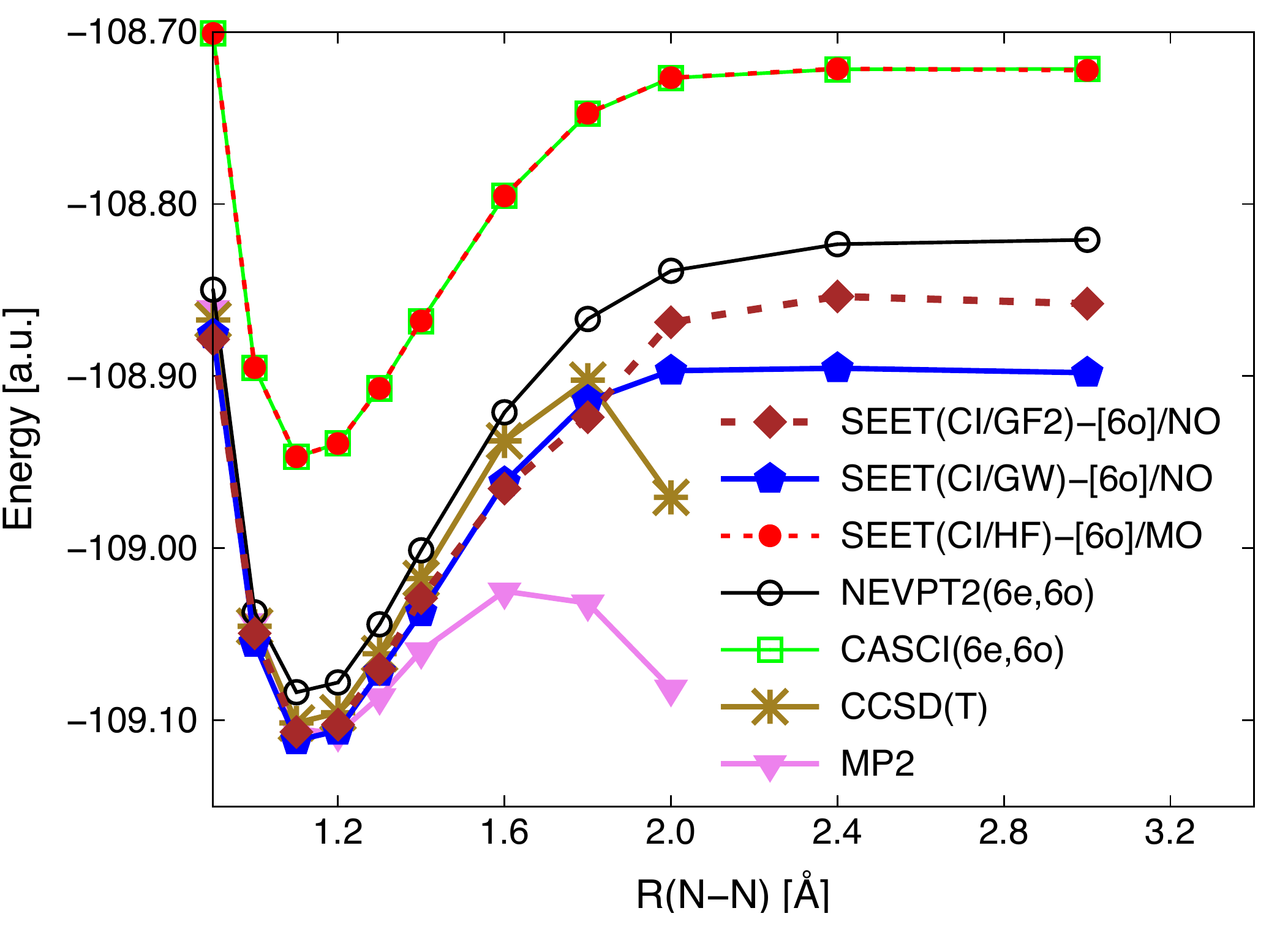}  \caption{Potential energy curves of N$_2$ in the 6-31G basis.}
  \label{fig:N2pec}
\end{figure}

In this section, we investigate the performance of SEET in combination with GW and GF2 for an N$_2$ molecule in the 6-31G basis. This system has 18 orbitals in total and six of them are strongly correlated for larger bond distances.
Stretching the triple bond of N$_2$ is a difficult test case for many quantum chemistry methods.
Potential energy curves evaluated with different variants of SEET and with standard wavefunction methods such as MP2, CCSD(T), CASCI, and NEVPT2 are shown in  Fig.~\ref{fig:N2pec}.

Standard single reference methods such as MP2 and CCSD(T) diverge once the bond is stretched.
CASCI(6e,6o) and NEVPT2(6e,6o), which capture the strong correlations arising among 6 orbitals when the triple bond is stretched well, produce correct dissociation curves.
The NEVPT2(6e,6o) energies are much lower than those of CASCI(6e,6o) due to the inclusion of weak correlations at the perturbative second order level between 6 strongly correlated orbitals and the remaining 12 weakly correlated orbitals. 
In CASCI(6e,6o), only correlations among the 6 strongly correlated orbitals are included, and the electronic effects among the remaining 12 orbitals are only illustrated at the HF level.
It is quite surprising that MP2 gives lower energies than NEVPT2. This behavior was previously observed in Ref.~\onlinecite{sokolov15}.

\begin{figure*} [bth]
  \includegraphics[width=17.0cm,height=5cm]{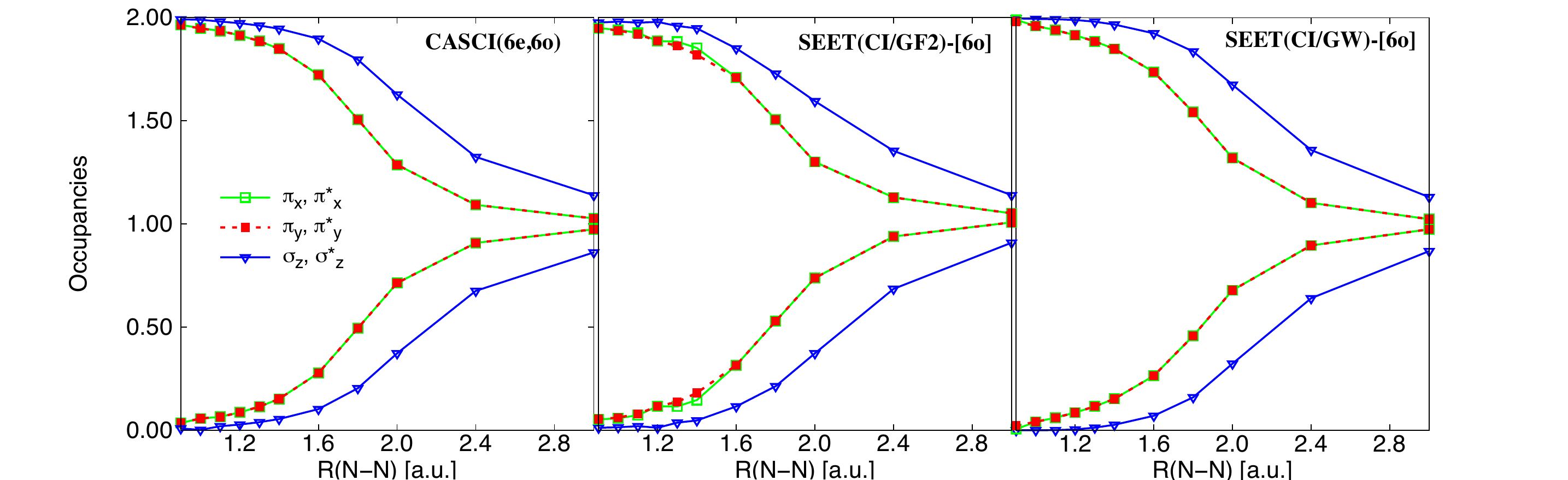}  \caption{Occupation numbers of 6 valence shell orbitals in N$_2$ ($\pi_x, \pi^*_x,\pi_y, \pi^*_y$, and $\sigma_z,\sigma^*_z$) obtained from CASCI(6e,6o) (left), SEET(CI/GF2)-[6o] (middle), and SEET(CI/GW)-[6o] (right) as a function of interatomic distance. Note that the $\pi$ orbitals are spatially degenerate.}
  \label{fig:N2occ}
\end{figure*}

To perform the SEET calculation, we use the following procedure.
First, SEET(FCI/HF)-[6o] using HF MOs is performed. Strongly correlated orbitals are identified as those 6 valence molecular orbitals with orbital energies closest to the Fermi level. These strongly correlated orbitals are delocalized and we do not perform any localization procedure.
On top of the SEET(FCI/HF)-[6o] reference, we carry out one iteration of GF2 or GW. The resulting natural orbitals as well as Green's functions and self-energies coming form this one iteration of GF2 or GW are then used to initialize  SEET(CI/GF2)-[6o] or SEET(CI/GW)-[6o] calculations.

The SEET(CI/HF)-[6o] curve is essentially identical to the CASCI(6e,6o) curve.
The weak correlations illustrated in GF2 and GW part of  SEET(CI/GW)-[6o] and SEET(CI/GF2)-[6o] contribute to a significant lowering of the energy when compared to SEET(CI/HF)-[6o].

Around the equilibrium geometry, both  SEET(CI/GW)-[6o] and SEET(CI/GF2)-[6o] yield energy curves close to  MP2 and CCSD(T).
Impressively, for stretched bond distances, unlike standard single reference methods such as MP2 and CCSD(T), SEET(CI/GF)-[6o] and SEET(CI/GW)-[6o] curves do not display any divergent behavior and remain parallel to those of CASCI(6e,6o) or NEVPT2(6e,6o).
This means that both SEET(CI/GF2)-[6o] and SEET(CI/GW)-[6o] properly illustrate the strong correlations necessary to qualitatively describe the N$_2$ dissociation. Additionally, they also include the weak correlations necessary for quantitative results. 
Note that SEET(CI/GF2)-[6o] displays a small kink near the geometry point where CCSD(T) diverges.

To further highlight that the correct description of correlation in SEET(CI/GF2) and SEET(CI/GW) is not a fortuitous coincidence, we present a comparison of occupation numbers.

For N$_2$ in the 6-31G basis, CASCI(6e,6o) yields almost identical occupation numbers to CASCI(10e,16o). The latter method correlates 16 orbitals at the FCI level while keeping the remaining two orbitals that have very low energy frozen. A detailed list of all the occupation numbers in CASCI(6e,6o), CASCI(10e,16o), SEET(CI/GF2), and SEET(CI/GW) is provided in the supplement.

For clarity, in Fig~\ref{fig:N2occ} only focuses on a comparison of the CASCI(6e,6o) occupation numbers for the six strongly correlated orbitals $\pi_x, \pi^*_x,\pi_y, \pi^*_y$ and $\sigma_z, \sigma^*_z$ to different variants of SEET.  It is evident that SEET(CI/GF2)-[6o] and SEET(CI/GW)-[6o] all yield occupation numbers very similar to those from CASCI(6e,6o).

\section{Relation of SEET(CI/GW) to GW+DMFT}\label{sec:SEET_vs_DMFT}
In this paper we discussed the theoretical framework and implementation for SEET with GW as a weakly correlated method and CI as a strongly correlated method. A related method, GW+DMFT,\cite{Biermann03,Biermann05,Biermann14} aims to describe the same class of physical problems where strongly and weakly correlated degrees of freedom are treated simultaneously. It is therefore instructive to summarize commonalities and differences.

The main difference between the two approaches is that SEET is built as an approximation to the Luttinger-Ward $\Phi$ functional, which is a functional of the Green's function and the bare interactions, whereas GW+DMFT is based on an approximation to the $\Psi$ functional, which is a functional of the Green's function and the screened interaction $W$. 
These distinct functional constructions lead to a fundamental difference in the type of impurity problem solved in SEET(CI/GW) and GW+DMFT. In SEET, the impurity problem is  defined by the bare interactions (transformed to an orbital basis), the non-interacting impurity Hamiltonian, and the impurity hybridization.
Consequently, impurity solvers such as configuration interaction\cite{Zgid11,Zgid12} or multi-orbital hybridization expansion quantum Monte Carlo\cite{Werner06b} can be employed.
In contrast, the impurity problem in GW+DMFT contains an additional dependence on the frequency-dependent interactions $W$.  Impurity solvers for GW+DMFT therefore have to be able to handle the frequency dependent interactions and work in an action representation.
Only few such solvers exist,\cite{Assaad07,Werner07,Werner10,Gull11} and they perform much better when hybridizations are diagonal and interactions are of the density-density type.\cite{Werner10} Consequently, current implementations of GW+DMFT either neglect the frequency dependence altogether or simplify the interaction structure of the impurity problem to simple density-density terms.

A second major difference is the choice of orbitals, which is always local in GW+DMFT, whereas SEET employs the freedom of choosing orbitals that minimize the hybridization to effectively decouple multiple impurity problems. 

Both of these methods are iterative self-consistent methods and are started by running an initial GW simulation. This initial solution may be far from the `true' self-consistent solution. How SEET(GW/CI) or GW+DMFT recover if the initial weak coupling answer is very different from the self-consistent fixed point is therefore an important question, especially since both SEET and GW+DMFT are highly non-linear, iterative procedures in which the final self-consistent result may depend strongly on the starting point.

SEET, when performed in a natural orbital basis (see the sections on H$_2$ and N$_2$), is capable of recovering from a qualitatively very different GW solution within the inner iteration of the self-consistency and the outer loop has a lesser influence.
When SEET is carried out using a localized orbital basis, the outer loop is very significant and is necessary to reach quantitative results and smooth potential energy surfaces.~\cite{Tran16} 

It is an interesting open question how the recovery from an inaccurate starting point is achieved in GW+DMFT. 
Here, the GW solution does not just enter the weakly correlated part of the system but also determines the impurity interactions via $W$. The self-consistent iteration therefore needs to adjust both the interactions and the hybridizations (rather than just the hybridizations). It may be for this reason that non-causal physics was reported for GW+DMFT,\cite{Haule17_h2} whereas no non-causal self-energies or Green's functions have so far been observed in SEET.

Finally, the possibility to use the generalized SEET framework  for treating intersecting subsets of orbitals leads to a systematically improvable method which provides an internal assessment of its accuracy. While a similar framework can be performed at the level of the $\Psi$ functional, this has not yet been implemented.

\section{Conclusions}\label{conclusions}
In summary, the major points observed during the numerical tests performed with SEET(CI/GW) are as follows. 

When SEET(CI/GW) is performed and all strongly correlated orbitals are placed in the impurity, a good agreement with standard quantum chemistry methods is achieved. We observe this in particular for the H$_2$ molecule in the TZ (Dunning) basis and for the N$_2$ molecule in the 6-31G basis. For N$_2$ both types of correlations, weak and strong, are recovered during the SEET procedure irrespective of whether GW or GF2 is used to treat weakly correlated electrons. In the equilibrium geometry, SEET agrees well with CCSD(T) and MP2. However, it avoids the divergence at the long distances, yielding curves parallel to NEVPT(6e,6o).

In the case of the hydrogen chains, we observe that when SEET-split is executed in a localized basis (SAO or LNO), both SEET with GW and GF2 lead to an overestimation of correlation energies for large interatomic distances. SEET(CI/GW) leads to potential energy curves more parallel to the FCI answer than SEET(CI/GF2). However, when LNOs are employed in a lager basis, SEET(CI/GW) yields overestimated electronic energies for all distances, not just for large distances.
The generalized version of the SEET functional with GF2, SEET(CI/GF2)-mix, can partially remedy this situation. In contrast, a similar improvement of SEET-split(CI/GW) by performing  SEET(CI/GW)-mix is unlikely, since the overestimation of SEET-split does not only come from using non-intersecting orbital subspaces but also from the GW approximation itself.

When the energy basis (NO) is used in conjunction with larger basis sets, we observe that for large interatomic distances SEET-split with both GF2 and GW leads to an underestimation of the total electronic energy. For small intermolecular distances SEET(CI/GW) and GW lead to an overestimation of the correlation energy.
As for the case of a localized basis, we observe that SEET-mix(CI/GF2) can help with recovering the correlations among strongly correlated orbitals and improve the results for the energies.
SEET-mix(CI/GW) results in potential energy curves that are more parallel to the exact answer. However, it cannot repair the energy overestimation due to GW itself.

%We notice that GW by itself leads to a large energy overestimation at small interatomic distances when larger basis sets are employed. We find this feature interesting and we believe that it should be examined carefully, since its presence may lead to a fortuitous and most likely unpredictable cancellation of errors when used with SEET or GW+DMFT.

For SEET(CI/GW) performed in a larger basis sets, additional studies are necessary not only to assess how correlations inside the strongly correlated space are recovered but also how they are captured in the space of weakly correlated orbitals. In particular, the errors arising due to the overestimation of correlation effects among the weakly correlated orbitals when a large orbital basis is used are presently unclear.
These questions require studies involving molecular cases where the strong correlations do not arise due to bond stretching but are intrinsic due to the presence of $d$- and $f$-electrons.

\section{Acknowledgements}

D.Z. and T.N.L. acknowledge funding by DOE Grant no. ER16391.
A.S. was supported by the NSF grant no. CHE-1453894. 
E. G. and J. L. are funded by the Simons collaboration on the many-electron problem.

\bibliographystyle{apsrev4-1}

\end{document}